\documentclass[letter]{jpsj2} 
\def\lsim{\mathstrut_{\displaystyle \sim}^{\displaystyle <}}
\def\gsim{\mathstrut_{\displaystyle \sim}^{\displaystyle >}}

\title{
Evidence of Strong Electron Correlations in $\gamma$-Iron
}

\author{
Yoshiro \textsc{Kakehashi}\thanks{E-mail address:
yok@sci.u-ryukyu.ac.jp, to be published in J. Phys. Soc. Jpn. {\bf 78}
No. 9 (2009).}, M. Atiqur R. \textsc{Patoary}, 
and Toshihito \textsc{Tamashiro}
}

\inst{
Department of Physics and Earth Sciences,
Faculty of Science, University of the Ryukyus, \\
1 Senbaru, Nishihara, Okinawa, 903-0213, Japan \\
}

\abst{
Single-particle excitation spectra of $\gamma$-Fe in the paramagnetic 
state have been investigated by means of the first-principles
dynamical coherent potential approximation theory which has recently
been developed. It is found that the central peak in the density of
states consisting of the t${}_{\rm 2g}$ bands is destroyed by electron
correlations, and the Mott-Hubbard type correlated bands appear.  
The results indicate that the $\gamma$-Fe can behave as
correlated electrons at high temperatures. 
}

\kword{single-particle excitation spectra, electron correlations, 
Mott-Hubbard band, $\gamma$-Fe, $\gamma$-Mn, Fe-pnictides, XPS, BIS}

\begin{document}
\maketitle

The 3$d$ transition metals are wellknown to behave as a typical itinerant
electron system in which the ground-state properties are explained well
by the band theory \cite{moruzzi95}.  
The cohesive properties such as lattice 
parameters and bulk moduli, and the ground-state magnetizations 
are in fact quantitatively explained by the density
functional theory for band calculations.  
On the other hand, excitations in these metals are often not explained by 
the band theory.  The magnetic properties of Fe, Co,
and Ni, for examples, show at finite temperatures the local moment 
behaviors as explained by the Heisenberg model \cite{kake04}.  
Excitation spectra of Ni observed by means of the X-ray photoelectron 
spectroscopy (XPS) are wellknown to show a
$d$-band narrowing and a satellite peak at 6 eV below the Fermi 
level \cite{himp79},
which can not be obtained by an independent-particle picture.
These results indicate that the effects of electron correlations 
 in transition metals strongly depends on the details of parameters
controlling the physical quantities. 
Small change in the key parameters might cause anomalous behaviors 
such as high-temperature superconductivity which has recently been 
found in the iron-arsenide system \cite{kami06}.
In order to clarify the characteristic features of
transition metal systems, one has to examine their electronic 
properties on the basis of realistic band theory which takes into 
account electron correlations.

In this letter, we present our numerical results of single-particle
excitation spectra for $\gamma$-Fe in the 
paramagnetic state, which are obtained by the first-principles dynamical
coherent potential approximation (CPA) theory, and demonstrate that
$\gamma$-Fe can be regarded as a strongly correlated 
electron system at high temperatures, though their ground-state 
properties are believed to be well explained by a band theory 
\cite{Fuj91,Kno00,Sjo02}. 

The first-principles dynamical CPA theory \cite{kake08} 
is the dynamical CPA \cite{kake92} combined 
with the first principle tight-binding (TB) linear muffin-tin orbital 
(LMTO) Hamiltonian \cite{ander94}.  
The former is a dynamical version of the single-site spin
fluctuation theory developed by Cyrot \cite{cyrot72}, Hubbard \cite{hub79}, 
and Hasegawa \cite{hase79}
since early in the 1970, and 
has recently been shown \cite{kake02-2} to be equivalent 
to the dynamical mean field theory (DMFT)~\cite{georges96}.  
The theory describes the electronic and magnetic properties at finite
temperatures efficiently taking into account the dynamical corrections 
to the spin and charge fluctuations.
Note that unlike the early quantum Monte-Carlo (QMC) calculations combined with 
the DMFT \cite{lich01} 
the present approach can treat the transverse spin
fluctuations for arbitrary $d$ electron number.

We adopt the TB-LMTO Hamiltonian $H_{0}$ plus the following intraatomic 
Coulomb interactions $H_{1}$ between $d$ electrons. 
\begin{eqnarray}
H_{1} = \sum_{i} 
\Big[ \sum_{m} U_{0} \, \hat{n}_{ilm \uparrow} \hat{n}_{ilm \downarrow} 
+ {\sum_{m > m^{\prime}}} 
(U_{1}-\frac{1}{2}J) \hat{n}_{ilm} \hat{n}_{ilm^{\prime}} -
{\sum_{m > m^{\prime}}} J   
\hat{\mbox{\boldmath$s$}}_{ilm} \cdot \hat{\mbox{\boldmath$s$}}_{ilm^{\prime}} 
\Big] \ . 
\label{h1}
\end{eqnarray}
Here $U_{0}$ ($U_{1}$) and $J$ are the intra-orbital (inter-orbital)
Coulomb interaction and the exchange interaction, respectively.
$\hat{n}_{ilm\sigma}$ is the number operator for electrons with orbital
$lm$ and spin $\sigma$ on site $i$.  
$\hat{n}_{ilm}$ ($\hat{\mbox{\boldmath$s$}}_{ilm}$) with $l=2$ is 
the charge (spin)
density operator for $d$ electrons on site $i$ and orbital $m$.

In the dynamical CPA \cite{kake08}, 
we transform the interacting Hamiltonian $H_{1}$
into a dynamical potential $v$ in the free energy adopting the
functional integral method, and expand the free energy with respect
to sites after having introduced a uniform medium, ({\it i.e.} 
a coherent potential) $\Sigma_{L\sigma}(i\omega_{n})$. 
Note that $L=(l,m)$, and $\omega_{n}$ denotes the Matsubara
frequency.  The first term in the expansion is the free energy for 
a uniform medium, 
$\tilde{\cal F}[\mbox{\boldmath$\Sigma$}]$.  
The second term is an impurity contribution to the
free energy.  The dynamical CPA neglects the higher-order terms, so that
the free energy per atom is given by
\begin{eqnarray}
{\mathcal F}_{\rm CPA} = \tilde{\mathcal F}[\mbox{\boldmath$\Sigma$}]
- \beta^{-1} {\rm ln} \, C \int d\mbox{\boldmath$\xi$} \,
{\rm e}^{\displaystyle -\beta E_{\rm eff}(\mbox{\boldmath$\xi$})} .
\label{fcpa2}
\end{eqnarray}
Here $\beta$ is the inverse temperature.  $C$ is a normalization
constant.  $\mbox{\boldmath$\xi$}$ denotes 
the static field variable on a site.  
$E_{\rm eff}(\mbox{\boldmath$\xi$})$ is an effective potential 
projected onto the static field $\mbox{\boldmath$\xi$}$.  
It consists of the static
term $E_{\rm st}(\mbox{\boldmath$\xi$})$ and the dynamical
correction term 
$E_{\rm dyn}(\mbox{\boldmath$\xi$})$.  The latter is given by a Gaussian
average of the determinant $D$ of the scattering matrix due to
dynamical potential as follows.
\begin{eqnarray}
{\rm e}^{\displaystyle -\beta E_{\rm dyn}(\mbox{\boldmath$\xi$})}
= \overline{D}=\overline{{\rm det}[1-(v-v_{0})\tilde{g}]} .
\label{edyn}
\end{eqnarray}
Here $(v)_{Ln\sigma L^{\prime}n^{\prime}\sigma^{\prime}}=
v_{L\sigma\sigma^{\prime}}(i\omega_{n}-i\omega_{n^{\prime}})
\delta_{LL^{\prime}}$ 
($(v_{0})_{Ln\sigma L^{\prime}n^{\prime}\sigma^{\prime}}
=v_{L\sigma\sigma^{\prime}}(0)\delta_{LL^{\prime}}\delta_{nn^{\prime}}$)
is the dynamical (static) potential, and 
$(\tilde{g})_{Ln\sigma L^{\prime}n^{\prime}\sigma^{\prime}}$
is the Green function in the static approximation.
The overline denotes the Gaussian average with respect to the field
variables. 

In order to treat the dynamical potential analytically, we expand the
dynamical part of the effective potential with respect to the 
frequency $\nu$ of 
$v_{L\sigma\sigma^{\prime}}(i\omega_{\nu})$, 
and neglect the mode-mode coupling terms.
This is called the harmonic approximation, and   
$\overline{D} \approx 1 + \sum_{\nu} (\overline{D}_{\nu}-1)$.
Here $D_{\nu}$ is a sub-matrix of $D$ in which the dynamical potential
$v$ has been replaced by its $\nu$ component only.

The effective medium ({\it i.e.}, the coherent potential) can be
determined by the stationary condition
$\delta \mathcal{F}_{\rm CPA}/\delta \Sigma = 0$.  
This yields the CPA equation as 
\begin{eqnarray}
\langle G_{L\sigma}(\mbox{\boldmath$\xi$}, i\omega_{n}) \rangle 
= F_{L\sigma}(i\omega_{n}) \ .
\label{dcpa3}
\end{eqnarray}
Note that $\langle \ \rangle$ at the l.h.s. 
is a classical average taken with respect to the
effective potential $E_{\rm eff}(\mbox{\boldmath$\xi$})$, 
$F_{L\sigma}(i\omega_{n}) = [(i\omega_{n} - \mbox{\boldmath$H$}_{0} 
- \mbox{\boldmath$\Sigma$})^{-1}]_{iL\sigma iL\sigma}$ 
is the coherent Green function, where 
$(\mbox{\boldmath$H$}_{0})_{iL\sigma jL^{\prime}\sigma}$ is the 
one-electron TB-LMTO Hamiltonian matrix, and 
$(\mbox{\boldmath$\Sigma$})_{iL\sigma jL^{\prime}\sigma} = 
\Sigma_{L\sigma}(i\omega_{n})\delta_{ij}\delta_{LL^{\prime}}$.
Furthermore, $G_{L\sigma}(\mbox{\boldmath$\xi$}, i\omega_{n})$
is the impurity Green function given by
\begin{eqnarray}
G_{L\sigma}(\mbox{\boldmath$\xi$}, i\omega_{n}) = 
\tilde{g}_{L\sigma L\sigma}(i\omega_{n}) + 
\frac{\displaystyle \sum_{\nu} 
\frac{\delta \overline{D}_{\nu}(\mbox{\boldmath$\xi$})}
{\displaystyle \kappa_{L\sigma}(i\omega_{n})\delta 
\Sigma_{L\sigma}(i\omega_{n})}}
{1+ \sum_{\nu} (\overline{D}_{\nu}(\mbox{\boldmath$\xi$})-1)} \ .
\label{gimp}
\end{eqnarray}
Here the first term at the r.h.s. is the Green function in the static
approximation. 
The second term is the dynamical corrections, and 
$\kappa_{L\sigma}(i\omega_{n})= 1 - F_{L\sigma}(i\omega_{n})^{-2}
\delta F_{L\sigma}(i\omega_{n})/\delta \Sigma_{L\sigma}(i\omega_{n})$.
\begin{figure}[b]
 \begin{center}
   \includegraphics{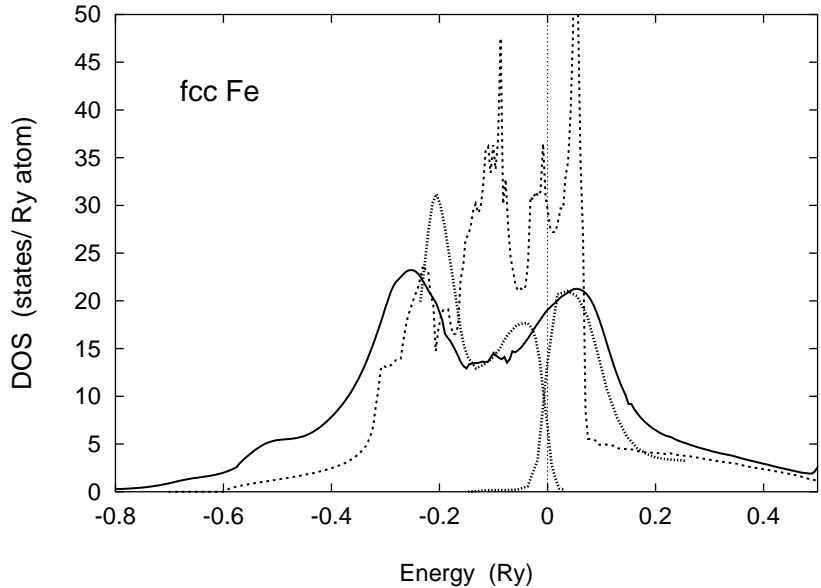}
 \end{center}
\caption{
Density of states (DOS) as the single-particle excitations for 
$\gamma$-Fe in the paramagnetic state calculated by the dynamical CPA 
(solid curve).
The DOS in the LDA is shown by dashed curve.  The BIS \cite{himp91} 
and XPS \cite{zhar97} data are shown by dotted curves. 
}
\label{fccfedos}
\end{figure}

Solving the CPA equation (4) self-consistently, we obtain the effective
medium $ \Sigma_{L\sigma}(i\omega_{n})$.  
The density of states (DOS) for the single-particle excitation spectrum 
is obtained by means of a numerical analytic continuation of the
coherent potential. 
In the numerical calculations, we expanded the dynamical correction term
$\overline{D}_{\nu}(\mbox{\boldmath$\xi$})$ in Eqs. (3) and (5)
with respect to the Coulomb interactions as 
$\overline{D}_{\nu}(\mbox{\boldmath$\xi$}) = 1 +
\overline{D}^{(1)}_{\nu}(\mbox{\boldmath$\xi$}) + 
\overline{D}^{(2)}_{\nu}(\mbox{\boldmath$\xi$}) + 
\overline{D}^{(3)}_{\nu}(\mbox{\boldmath$\xi$}) + 
\overline{D}^{(4)}_{\nu}(\mbox{\boldmath$\xi$}) + \cdots$,
and calculated the r.h.s. up to the second order exactly as has been
made in our previous calculations \cite{kake08}.
In addition to the second-order terms, the third and fourth order terms 
are taken into account approximately in the present calculations
by using an asymptotic approximation \cite{kake92}.
We have obtained the TB-LMTO Hamiltonian at the observed lattice 
constant 6.928 a.u. at 1440 K using the local density approximation
(LDA) in the density functional theory.
The average Coulomb and exchange energy parameters 
($\overline{U}$ and $J$) for Fe are
taken from the LDA+$U$ band calculations \cite{anisimov97} ;
$(\overline{U},J)= (0.169,0.066)$.
$U_{0}$ and $U_{1}$ are obtained from the relations 
$U_{0} = \overline{U}+8J/5$ and $U_{1} = \overline{U} - 2J/5$.

We calculated the DOS for $\gamma$-Fe in the
paramagnetic state at high temperatures ($T=2000$K) where the present
theory works best.  The results are presented in
Fig. 1.  The fcc DOS in the LDA is characterized by the main peak near
the top of $d$ bands, the central peak around $\omega = -0.1$ Ry, and the
third peak near the Fermi level.  The first two peaks originate in
the t${}_{\rm 2g}$ bands and the third one is due to the 
e${}_{\rm g}$ bands. 
In the dynamical CPA calculations, 
the central peak is destroyed by electron correlations 
and the DOS shows the two peaks.
\begin{figure}[t]
 \begin{center}
   \includegraphics{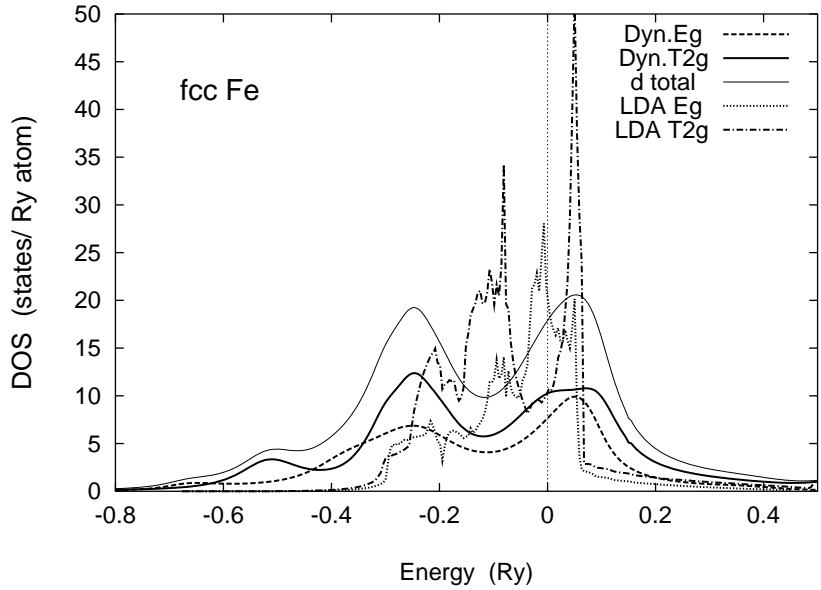}
 \end{center}
\caption{
Partial $d$ DOS for e${}_{\rm g}$ (dashed curve) and t${}_{\rm 2g}$ 
(solid curve) electrons.  The total $d$ DOS are shown by thin solid 
curves. The partial DOS in the LDA are shown by dotted curve 
(e${}_{\rm g}$) and dot-dashed curve (t${}_{\rm 2g}$). 
}
\label{fccfelmdos}
\end{figure}
The two peak structure is more explicitly found in the partial $d$ 
DOS for e${}_{\rm g}$ and t${}_{\rm 2g}$ orbitals as shown in Fig. 2.
Both local DOS show a dip at $\omega=-0.1$ Ry.  Especially the central
peak of the t${}_{\rm 2g}$ bands in the LDA disappears due to electron
correlations and the spectral weight moves to the lower and higher
energy regions ({\it i.e.}, $\omega \, \lsim \, -0.3$ Ry and 
$\omega\, \gsim \,0.1$ Ry).  
These changes in the DOS are caused by a strong scattering peak of $-{\rm
Im} \Sigma_{L\sigma}(\omega+i\delta)$ around $\omega=-0.1$ Ry where the
central peak of the t${}_{\rm 2g}$ band in the LDA is located, and 
also by a change of the energy shift 
Re$ \Sigma_{L\sigma}(\omega+i\delta)$ in sign; 
Re$ \Sigma_{L\sigma}(\omega+i\delta) < 0$ for $\omega < -0.1$ Ry and 
Re$ \Sigma_{L\sigma}(\omega+i\delta) > 0$ for $\omega > -0.1$ Ry.
The two peaks at $\omega=-0.25$ and $\omega=0.05$ Ry in the DOS are
therefore considered to be the lower and upper Mott-Hubbard bands as 
found in the half-filled Hubbard model.

There is no experimental data for the bulk $\gamma$-Fe at high 
temperatures as far as we know.  The BIS experimental data for 
8 fcc Fe monolayers on Cu (100) surface at room
temperature \cite{himp91} are consistent with our results as shown in
Fig. 1.  The photoemission data \cite{zhar97} 
for 5 fcc Fe monolayers on Cu (100) are also shown in the figure.  
The peak at $\omega=-0.205$ Ry is usually interpreted to
be due to the emission from the bulk Cu substrate.  
Another interpretation is that both the lower Hubbard bands for Fe and 
the Cu-substrate bands are superposed.  Assuming the latter, the present
results are consistent with the experimental data.

We have also performed the DOS calculations for $\gamma$-Mn to make sure
the formation of the Mott-Hubbard bands in the fcc transition metals.  
We adopted the average Coulomb and exchange parameters 
$(\overline{U},J)= (0.192, 0.061)$, which are taken from the LDA + $U$
calculations with use of the Hartree-Fock-Slater potentials \cite{banddyo89}
and from the atomic calculations \cite{mann67}, respectively.
We find again the two-peak structure in the DOS as shown in Fig. 3 
when the dynamical CPA is applied.  
The energy difference between the lower peak and the upper one is
larger than that of $\gamma$-Fe by about 0.1 Ry and the valley between
the peaks becomes deeper by a factor of two.
The present results are consistent with those calculated by the QMC+DMFT 
with use of the Hamiltonian without transverse spin fluctuations
\cite{bier01}.
\begin{figure}[tb]
 \begin{center}
   \includegraphics{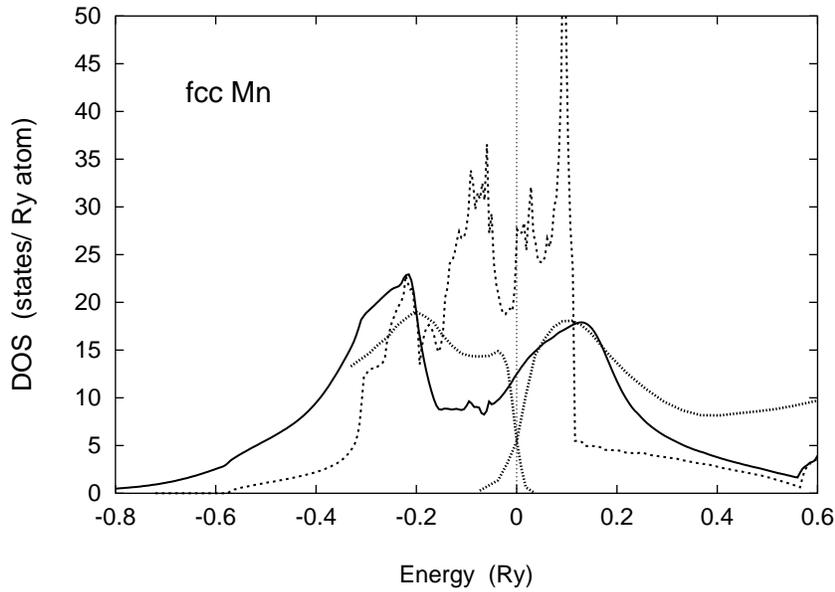}
 \end{center}
\caption{
The DOS for $\gamma$-Mn in the paramagnetic state (solid curve).
The DOS in the LDA is shown by dashed curve.  The XPS \cite{bier01} 
and BIS \cite{speier84} data are shown by dotted curves. 
}
\label{fccmndos}
\end{figure}

There is no XPS experimental data for the bulk $\gamma$-Mn.  
But, the photoemission data \cite{bier01} for the 20 monolayer fcc Mn on
the Cu${}_{3}$Au (100) are consistent with the results of calculations 
as shown in Fig. 3.
The high-energy peak at $\omega=0.1$ Ry
may be justified by the BIS data \cite{speier84}, although the data
are taken for $\alpha$-Mn at room temperature.  

It should be noted that in the case of Ni the central peak of the 
t${}_{\rm 2g}$ bands in the LDA shifts to $\omega=-0.15$ Ry.  
Furthermore, the scattering magnitude $-{\rm
Im}\Sigma_{L\sigma}(\omega+i\delta)$ is weakened by a factor of ten or
more as compared with the case of $\gamma$-Fe, 
and it is enhanced around $\omega=-0.45$ Ry instead of $\omega=-0.15$ Ry.
Moreover the energy shift 
Re$ \Sigma_{L\sigma}(\omega+i\delta) < 0$ for $\omega > -0.45$ Ry and 
Re$ \Sigma_{L\sigma}(\omega+i\delta) > 0$ for $\omega < -0.45$ Ry.
Therefore the valley found in the $\gamma$-Fe at $\omega=-0.1$ Ry 
disappears in the case of Ni.  Instead, a small satellite peak appears 
around $\omega=-0.50$ Ry (6 eV satellite \cite{himp79}).

We note that the Mott-Hubbard bands have recently been found in 
the Fe-based compound LaFeAsO.  
There Fe atoms form a two-dimensional square lattice.
The DOS \cite{haule08} 
calculated by the DMFT clearly show a two-peak structure 
with a deep depression around $\omega=-0.1$ Ry, 
which is essentially the same as the present results of calculations
though the LaFeAsO system seems to be considerably stronger than 
the $\gamma$-Fe in Coulomb interaction strength.

In summary, we have calculated the DOS of single-particle excitation
spectra for $\gamma$-Fe in the paramagnetic state on the basis of the
first-principles dynamical CPA.  We found that 
the $\gamma$-Fe shows the strong electron correlation effects on the 
the single-particle excitations at high temperature region,
{\it i.e.}, the disappearance
of the central peak of the t${}_{\rm 2g}$ bands due to electron
correlations and a formation of the Mott-Hubbard type bands. 
The result is quite different from the itinerant electron picture of 
$\gamma$-Fe obtained by the ground-state band calculations 
\cite{Fuj91,Kno00,Sjo02}. 
Systematic investigations of the bulk fcc transition metals at high
temperature region with use of the photoelectron spectroscopy are highly
desired to verify the present results.

\section*{Acknowledgement}
This work was supported by Grant-in-Aid for Scientific Research (19540408). 
Numerical calculations have been partly carried out with use of 
the Hitachi SR11000 in the Supercomputer Center, Institute of 
Solid State Physics, University of Tokyo.

\end{document}